\documentclass[11pt]{article} 

\usepackage{setspace}
\usepackage{tikz}

\usepackage{graphicx,verbatim}
\usepackage{amsfonts,amsmath}
\usepackage{psfrag}
\usepackage{accents}
\usepackage{mathrsfs}
\usepackage[colorinlistoftodos]{todonotes}
\usepackage{epstopdf}

\def\be{\begin{equation}}
\def\ee{\end{equation}}
\def\ba{\begin{eqnarray}}
\def\ea{\end{eqnarray}}
\def\bsP{P_{\!a}^{\rm BS}}
\def\admP{P_{\!a}^{\rm ADM}}
\def\Rh{\hat{\mathcal{R}}}
\def\psRh{\boldsymbol{\Gamma}_{\!\scriptscriptstyle{\cal {\hat R}}}}

\def\={\hat{=}}

\def\tl{\tilde}
\def\h{\hat}
\def\f{\frac}

\def\rmd{\mathrm{d}}

\def\b{\bar}
\def\qo{\mathring{q}}
\def\go{\mathring{g}}
\def\Do{\mathring{D}}
\def\no{\mathring{n}}

\def\ro{\mathring{r}}
\def\rmd{\mathrm{d}}

\def\Lie{\mathcal{L}}
\def\E{\mathcal{E}}
\def\S{\mathcal{S}}
\def\R{\mathcal{R}}
\def\N{\mathcal{N}}
\def\No{\N_\circ}
\def\rad{\mathfrak{r}}

\newcommand*{\scri}{\ensuremath{\mathscr{I}}} 
\newcommand*{\scrip}{\ensuremath{\mathscr{I}^{+}}} 
 
\newcommand*{\scripm}{\ensuremath{\mathscr{I}^{\pm}}} 
\def\inot{i^\circ}

\def\no{\mathring{n}}

\begin{document}

\title{\bf{The Operational Meaning of Total Energy\\
 of  Isolated Systems in General Relativity}}

\author{Abhay Ashtekar$^{1}$, Simone Speziale$^{2}$ 
\smallskip \\ 
$^{1}$ \small{ Physics Department, Penn State, University Park, PA 16802, U.S.A.,} \\
\small{and Perimeter Institute for Theoretical Physics, 31 Caroline St N, Waterloo, ON N2L 2Y5, Canada}\\
$^{2}$ \small{ Aix Marseille Univ., Univ. de Toulon, CNRS, CPT, UMR 7332, 13288 Marseille, France}}

\maketitle

\begin{abstract}
We present thought experiments to measure the Arnowitt-Deser-Misner $E_{\rm ADM}$ and Bondi-Sachs energy $E_{\rm  BS}$ of isolated systems in general relativity. The expression of $E_{\rm BS}$ used in the protocol is likely to have other applications. In particular, it is well-suited to to be promoted to an operator in non-perturbative loop quantum gravity.
\\
\medskip
\\
\emph{Keywords:} Isolated gravitating systems, Asymptotic flatness, Observables
\end{abstract}



\section{Introduction}
\label{s1}

It is a pleasure to write this article in honor of Jorge Pullin on the occasion of his 60th birthday. Jorge has made important contributions to loop quantum gravity and classical general relativity over the years and, rather astonishingly, he is currently engaged in experimental work at the LIGO Livingston facility. We thought it would be appropriate to honor his extraordinary breadth of interests through an article that touches on all three areas --albeit only through \emph{thought} experiments in the third area since, unlike Jorge, we cannot come even close to participating in actual hands-on experiments! \vskip0.1cm 

Total energy $E_{\rm total}$ is perhaps the most basic observable of an isolated system in general relativity. However, the notion is surprisingly subtle for the following reasons:\vskip0.2cm

\indent (i) In Newtonian gravity the analogous observable is the total mass which can be obtained by integrating the matter density over sources, or, by computing the flux of the `electric field' --gradient of the Newtonian potential-- across any finite 2-sphere that surrounds matter sources. In general relativity, this is not possible. For, in generic situations, there is no invariantly defined scalar potential, nor a 2-form that one can integrate on a finite 2-sphere surrounding sources to obtain a quantity with dimensions of energy that is independent of the choice of the 2-sphere. This feature is a consequence of a fundamental conceptual obstruction: in general relativity, the gravitational field also contributes to the total energy momentum, the `gravitational charge'. A source-free Maxwell field in Minkowski space-time, for example, does not contribute to the electric charge, while a source-free solution to Einstein's equations does carry energy-momentum (even when space-time is topologically $\mathbb{R}^4$). In the gauge theory language, general relativity is analogous to a \emph{non-Abelian} Yang-Mills theory, where the total charge has to be defined by an integral at infinity. Similarly, in general relativity, total energy \emph{can be only defined using 2-sphere integrals in asymptotically flat regions} of space-time.
\vskip0.15cm
\indent (ii) In Newtonian gravity, the mass of an astrophysical body can be determined operationally by observing the orbits of test bodies moving  around it. One can extend this idea to general relativity if the central body is stationary. However, in general asymptotically flat situations, this is not possible. Now mass becomes a genuinely \emph{global} notion in that it requires an examination of the behavior of test particles distributed across \emph{an entire 2-sphere} 
in the asymptotic region. That the notion of total energy in general relativity is genuinely global\, `in angles'\, is brought to forefront by a theorem due to Carlotto and Schoen \cite{carlotto-schoen,schoen,pc-bourbaki}. Using gluing techniques, they have shown that the initial value equations of general relativity admit asymptotically flat, regular solutions on a 3-manifold $\Sigma$  with positive Arnowitt-Deser-Misner (ADM) energy that have an astonishing property: Outside a cone $\mathcal{C}$ extending to infinity, the Cauchy data on $\Sigma$ are  isomorphic to that on a Cauchy surface in Minkowski space. (In the literature, the ADM energy is sometimes referred to as ADM mass.) $\Sigma$ can be taken to be topologically $\mathbb{R}^3$ and the angular span of $\mathcal{C}$ can be very small! In the space-time obtained by evolving such data, the 4-metric $g_{ab}$ is \emph{flat} in the domain $\mathcal{D}$ of dependence of the complement of\, $\mathcal{C}$\, in\, $\Sigma$. Test particles in this region will not experience any physical acceleration nor tidal forces! Therefore observations of the orbits of test particles in $\mathcal{D}$ will yield no information about the ADM mass of these space-times. As a consequence, once the requirement of stationarity is dropped, experiments --\emph{even thought experiments}-- aimed at measuring the total energy (or just checking if it is non-zero), require a \emph{full 2-sphere worth} of test particles in the asymptotic region, in striking contrast to Newtonian theory. \vskip0.15cm  
\indent (iii) The standard definition of the ADM mass is not arrived at by invoking any operational protocols or thought experiments.  Rather, it is based on Hamiltonian methods and uses asymptotic symmetries \cite{adm,rt}. More precisely, one constructs the phase space of asymptotically flat initial data on a 3-manifold $\Sigma$ and \emph{defines} components of the ADM 4-momentum $\admP$ as Hamiltonians generating \emph{asymptotic} space-time-translations at spatial infinity. The ADM energy $E_{\rm ADM}$ is the time component of $\admP$ in the asymptotic rest frame defined by $\Sigma$. This mathematical procedure leads to the conclusion that the ADM energy depends \emph{only} on the configuration variable (the positive definite 3-metric $q_{ab}$ on the 3-manifold $\Sigma$) and is independent of momentum ($p^{ab}$, encoded in the extrinsic curvature). This does not happen in \emph{any} non-gravitational field theory of physical interest!

Furthermore, even in general relativity, this result seems surprising from the perspective of stationary space-times in which there is an exact time-translation symmetry. For, even before the advent of the ADM framework, it was known that in stationary space-times one can define the Komar charge \cite{komar} which also carries the interpretation of total the energy, but which is constructed from the norm of the Killing field that resides in the `time-time component' of the 4-metric $g_{ab}$, rather than in the spatial metric $q_{ab}$. Indeed, this fact had led some leading researchers to doubt the correctness of the ADM energy.%
{\footnote{In particular, this concern was explicitly voiced by Arthur Komar in the early 1980s during a discussion on asymptotic flatness with AA and Peter Bergman.}}\,
It took a careful analysis \cite{beig,aaam-conserved,aaam3+1} to show that the two apparently distinct expressions in fact agree in stationary space-times that are asymptotically flat at spatial infinity.

Finally, the expression of the full ADM 4-momentum $\admP$ involves \emph{only} the gravitational variables --the positive definite 3-metric $q_{ab}$ and its conjugate momentum $p^{ab}$. Matter fields do \emph{not feature} in the expression of $\admP$ even when their support extends all the way to infinity, as is often the case for the electromagnetic field in the Einstein-Maxwell theory. Yet $\admP$  represents the \emph{total} 4-momentum of the isolated system, \emph{including contribution from sources.} \vskip0.15cm
\indent (iv) In contrast to $\admP$, the  Bondi-Sachs 4-momentum $\bsP$ at null infinity $\scrip$ is dynamical in that it changes in response to the radiation escaping through $\scrip$ and thus has richer content that $\admP$. It was originally identified  by imposing suitable boundary conditions as one recedes from sources in null directions, examining asymptotic field equations, and requiring that the radiated energy should be positive \cite{bondi,sachs}. In contrast to the ADM 4-momentum $\admP$, the Bondi-Sachs 4-momentum $\bsP$ does \emph{not} have the interpretation of a Hamiltonian in any phase space framework if the cross-section on which it is defined has non-zero news.
\footnote{The Bondi-Sachs energy-momentum \emph{flux} $F_a [\Rh]$ through a 3-d region $\Rh$ of $\scrip$ does arise as the Hamiltonian generating Bondi-Metzner-Sachs (BMS) translations on the phase space $\psRh$ consisting of radiative modes that reside in the region $\Rh$\,\,\cite{aams,aass2}. The Bondi-Sachs 4-momentum aspect\, (a 2-form on $\scrip$)\, can then be obtained by a systematic `integration' of the flux aspect\, (a 3-form on $\scrip$). But the resulting charge, $\bsP$ is not a generator of the BMS symmetry in any Hamiltonian framework that allows gravitational waves, i.e., in which $\scrip$ is a `leaky boundary'.}
\vskip0.1cm

These considerations spell out the mathematical reasoning that led to the definition of 4-momenta $\admP$ and $\bsP$ of isolated systems in general relativity. The Hamiltonian framework underlying $\admP$ and the balance laws that are used to arrive at $\bsP$ are powerful guidelines. Positivity of the resulting ADM and Bondi-Sachs energy are deep results \cite{sy,ew,ghmp,sy2,orpt} that provide a strong support for the final expressions of $\admP$ and $\bsP$. Yet, the arguments are rather indirect and the expressions are rather mysterious from a physical standpoint. In the Newtonian limit, the ADM energy in the rest frame of the system does yield the Newtonian mass. But, as we see from the Carlotto-Schoen construction, highly non-trivial possibilities arise in full general relativity. Is there a more direct physical sense in which the mass of a general isolated system that the tests bodies in the asymptotic regions respond to is the one provided by ADM 4-momentum? The expression of Bondi-Sachs 4-momentum is even further removed from direct physical considerations. As we discuss below, in the Newman-Penrose framework it is a specific combination of the Coulombic field (the Weyl tensor component $\Psi_2^\circ$) that\, `falls off as $1/r^3$'\, and radiative fields (the Bondi news $\dot\sigma^\circ$ and the asymptotic shear  $\sigma^\circ$) that\, `fall-off as $1/r$'. A priori it is rather mysterious why this precise combination is necessary to capture the notion of the total energy of the isolated system at a given retarded instant of time. In particular, we cannot make appeal to the Newtonian limit to argue for these choices since there are no gravitational waves in Newtonian theory.

The purpose of this paper is to suggest thought experiments to measure the ADM and Bondi-Sachs energies. In each case we will first recast the original expressions of these observables to a form that makes their physical content more transparent. The experiments will provide a precise sense in which they carry the connotations that we physically expect of total energy, given the well-established features of general relativity. This exercise provides some surprising insights and will have interesting applications as well. 

In section \ref{s2} we consider space-times that are asymptotically flat at spatial infinity. We will first recall that the standard expression of ADM energy can be recasts in terms of curvature in a number of ways \cite{aaam3+1}.\, One of these equivalent expressions involves the leading order part of the 4-dimensional Riemann tensor that dictates the  
tidal acceleration between test particles in the asymptotic region. Therefore, it is possible to introduce a thought experiment to measure the ADM energy (in any given asymptotic rest frame) by observing the \emph{relative} acceleration between test particles distributed on a 2-sphere with a large area-radius $R$,\, and a nearby one with radius $R -\delta$. This experiment provides an operational meaning to the ADM energy by tying it with tidal acceleration, the hallmark of the physical effect of gravity in the framework of general relativity. And this thought experiment measures the ADM energy for \emph{generic} gravitational systems, including the exotic ones represented by initial data of Carlotto-Schoen type. Furthermore the thought experiment in terms of tidal accelerations can also be used directly in Newtonian gravity to measure mass \cite{aasb}. In the passage from Newtonian gravity to general relativity the very meaning of `gravity' is profoundly altered. But the notion of tidal acceleration carries over. Therefore, one can make the case that the essence of the physical effect of the total energy of an isolated system lies in the tidal accelerations it causes, irrespective of whether one uses Newtonian gravity or general relativity. It is then natural to devise thought experiments to measure the gravitational mass by focusing on tidal accelerations.

In section \ref{s3} we consider space-times that are asymptotically flat at null infinity. In absence of gravitational waves, as one would expect, one can use the same thought experiment as in section \ref{s2}\, \cite{aasb}. However, the strategy is not viable in presence of gravitational radiation. In retrospect, this is not that surprising because the Newtonian intuition fails in the description of processes in which gravitational waves carry away energy. While the Bondi-Sachs expression provides a specific prescription for including contributions from gravitational waves, as we already noted, \emph{physically} it is rather obscure as to why the specific term that is includes is the appropriate one. We will first recast that expression using properties of ingoing null geodesics. The new expression will show that in all space-times that are asymptotically flat at null infinity, the total Bondi-Sachs energy is a measure of the rate at which the ingoing light front emanated by a 2-sphere contracts in the asymptotic region. Therefore one can devise a thought experiment to measure the Bondi-Sachs energy $E_{\rm BS}$ in terms of the asymptotic properties of null geodesics. Note that the experiment can be used also in situations in which \emph{$E_{\rm BS}$ is time-changing due to outgoing gravitational waves.} In retrospect, it is perhaps not surprising that one has to use properties of null geodesics for Bondi-energy in place of time-like geodesics used for ADM energy. We focus on these two notions of energy because they are 
by far the most important ones, widely used in the community. Other notions (see, e.g., \cite{szabados}) are neither as compelling nor have they proved to be as useful.

In section \ref{s4} we summarize the results and suggest some future applications.  In a nutshell, the standard expressions of $E_{\rm ADM}$ and $E_{\rm BS}$ were arrived at using  well-motivated mathematical considerations. However, one also expects the presence of a gravitational mass to manifest itself via  physical effects on test bodies, and a priori it is far from clear how, and indeed why, such effects are correctly encapsulated in these expressions. Therefore it is interesting that, using the notion of asymptotic flatness in the respective regimes, the expressions can be recast to a form which brings out a precise sense in which they have the correct physical connotations, given our understanding of the interplay between physics and geometry encoded in general relativity. Specifically, these expressions suggest thought experiments to measure $E_{\rm ADM}$ and $E_{\rm BS}$ in \emph{generic} asymptotically flat space-times. The specific expressions will also be useful in other contexts in both classical and quantum gravity.

Our conventions are as follows. Physical space-times are denoted by $(M, g_{ab})$ where $g_{ab}$ has signature -,+,+,+. The torsion-free derivative operator compatible with $g_{ab}$ is denoted by $\nabla$ and its curvature tensors are defined via:  $2\nabla_{[a}\nabla_{b]} v_c = R_{abc}{}^d v_d$,\, $R_{ac} = R_{abc}{}^b$, and $R=g^{ab} R_{ab}$.  The decomposition of the Riemann tensor into trace free and trace parts is given by: $R_{abcd} = C_{abcd} + g_{a[c} S_{d]b} - g_{b[c} S_{d]a}$, where $C_{abcd}$ is the Weyl tensor and $S_{ab} = R_{ab} - \f{1}{6} R\, g_{bc}$ the Schouten tensor constructed from the Ricci tensor $R_{ab}$.

\section{The ADM Energy}
\label{s2}

In this section we will begin by recasting the standard expression of the ADM energy $E_{\rm ADM}$ in a form that refers to the 4-d Riemann tensor rather than the 3-d metric. We will then use this new form to introduce a thought experiment involving tidal accelerations to measure $E_{\rm ADM}$. The thought experiment is motivated by Newtonian considerations but encompasses \emph{all} gravitational fields of general relativity, including those that have no Newtonian analogs. By carrying it out in different asymptotic rest-frames one recovers the full information in $\admP$.

\subsection{Preliminaries}
\label{s2.1}

The ADM 4-momentum is defined in the setting of 3+1 Hamiltonian framework. Let us fix a 3-manifold $\Sigma$ which is topologically $\mathbb{R}^3$ outside a compact set so that it has only one asymptotic end.  The fields of interest are a positive-definite 3-metric $q_{ab}$, its conjugate momentum $p^{ab}$, and the matter density $\rho$ and current $j^a$. 
From the perspective of the 4-d metric $g_{ab}$ obtained by evolution, $p^{ab}$ is related to the extrinsic curvature $k^{ab}$ via {\smash{$p^{ab}=\f{\sqrt{q}}{G} (k^{ab} - k q^{ab})$}} and the matter density and current are related to the stress-energy tensor via $\rho = T_{ab} \tau^a \tau^b$ and $j^a = T_{bc} \tau^b q^{ac}$, where $\tau^a$ is the unit time-like normal to $\Sigma$ with respect to the 4-metric $g_{ab}$. The initial data $(q_{ab}, k_{ab}, \rho, j^a)$ are subject to the scalar and vector constraints of general relativity: 
\be \R - k_{ab} k^{ab} + k^2 = 16\pi G\, \rho, \quad {\rm and} \quad D_a(k^{ab} - k q^{ab}) = 8\pi G\, j^b. \ee
Here $D$ and $\R$ are the (torsion-free) derivative operator and the scalar curvature defined by $q_{ab}$. In the positive energy theorems one assumes the energy condition $\rho \ge (j_a j^a)^{\f{1}{2}}$.

We are interested in the initial data that are asymptotically flat at spatial infinity. Thus, we require that, outside a compact set, $\Sigma$ admits a flat metric $\qo_{ab}$ and impose fall-off conditions on components of various fields in the Cartesian coordinates $x^{\b{k}}$ of $\qo_{ab}$: \vskip0.1cm
\noindent(i)  $(q_{\b{a}\b{b}}\, -\, \qo_{\b{a} \b{b}})\, =\, O(\f{1}{r});\quad \Do_{\b{c}}\,q_{\b{a}\b{b}}\, =\, O(\f{1}{r^2})$;\quad and, \quad $\Do_{\b{c}}\Do_{\b{d}}\,q_{\b{a}\b{b}}\, =\, O(\f{1}{r^3})$;\\
(ii) $k_{\b{a}\b{b}} = O(\f{1}{r^2})$; \quad and \quad $\Do_{\b{c}}\,k_{\b{a}\b{b}} = O(\f{1}{r^3})$;\\
(iii) $\rho, \,\,j_{\b{a}}$,\,\, and the space-space components \,\, $t_{\b{a} \b{b}}$ of $T_{ab}$ are all $O(\f{1}{r^4})$
\vskip0.1cm 
\noindent as $r \to \infty$. Here $r$ is the radial coordinate defined by $x^{\b{k}}$ and $\Do$ the (torsion-free) derivative operator compatible with $\qo_{ab}$.  The fall-off condition on the stress-energy tensor are motivated by the asymptotic behavior of physically interesting Maxwell fields at spatial infinity. 

Given such an initial data, the ADM energy $E_{\rm ADM} = - \tau^a \admP$ is defined by 
\be \label{ADMenergy} E_{\rm ADM} = \f{1}{16\pi G} \, \lim_{\ro\to \infty}\,\oint_{r=r_\circ} (\Do_c\,q_{bc} - \Do_b\, q_{ac})\, \qo^{ac}\,{\ro}^b \rmd^2V \ee
where ${\ro}^b = \qo^{ab} \Do_a r$ is the unit normal to the $r=r_\circ$\, 2-spheres w.r.t. $\qo_{ab}$. Thanks to our conditions on the initial data, the integral is well-defined. Furthermore (although it is far from being clear from its expression), it is positive and vanishes if and only of the initial data corresponds to Minkowski space-time\,\cite{sy,ew}. 

\subsection{Expressing the ADM Energy in terms of Curvature}
\label{s2.2}

To recast \eqref{ADMenergy} in terms of the Riemann tensor $R_{abcd}$ of the 4-d solution $g_{ab}$ we will proceed in three steps. In the first, we will recast it using of the Ricci curvature $\R_{ab}$ of the physical 3-metric $q_{ab}$,\, in the second, in terms of the Weyl curvature $C_{abcd}$ of the 4-metric $g_{ab}$,\, and, in the third, using the full  Riemann tensor $R_{abcd}$. These expressions will refer only to the asymptotic behavior of $r$ and not to other features of the fiducial metric $\qo_{ab}$, such as its derivative operator $\Do$. 

To carry out the first step, one uses the following fact. The difference between the derivative operators $D$ and $\Do$ is captured in a tensor field $C_{ab}{}^c$:
\ba \label{C} &(D_a - \Do_a) k_b = C_{ab}{}^c k_c\quad \forall\, k_c, \qquad {\rm where} \nonumber\\
&C_{ab}{}^c = - \f{1}{2}\, q^{cm}\, \big(\Do_a q_{bm} + \Do_b q_{am}  - \Do_m q_{ab}\big) \ea
and, since $\qo_{ab}$ is flat, the Ricci tensor $\R_{ab}$ of $q_{ab}$ is given by
\be \label{Ricci} \R_{ac} = 2 \big(\Do_{[a}\, C_{b] c}{}^b  + C_{c[a}^d\, C_{b] d}{}^b \big). \ee
One then substitutes the expression (\ref{C}) of $C_{ab}{}^c$ in (\ref{Ricci}) and uses the fall off condition (i) in the definition of asymptotic flatness, and the fact that \smash{$r\,\Do_a \Do_b  r = \big(\qo_{ab} - \Do_a r \Do_b r)  =: \mathring{\tilde{q}}_{ab}$}  is the metric induced by $\qo_{ab}$ on the \smash{$r=r_\circ$}\,\, 2-spheres. Then, in terms of the radial-radial component of $\R_{ab}$, we have: 
\be \label{ricci-exp1} -\, \,\oint_{r=r_\circ}\!\!\!\! r\, \ro^a \ro^b\, \R_{ab}\, \rmd^2V\, =\, {\textstyle{\f{1}{2}}}\,  \lim_{r_\circ\to \infty}\,\oint_{r=r_\circ}\!\!\! (\Do_c\,q_{bc} - \Do_b\, q_{ac})\, \qo^{ac} \tau^b \rmd^2V\, , \ee
where $\ro^a$ denotes the unit normal to the $r=r_\circ$ 2-spheres w.r.t. $\qo_{ab}$. In arriving at (\ref{ricci-exp1}), one has to perform integration by parts and use Stokes' theorem to set certain terms to zero.%
\footnote{More precisely, one uses the fact that the 2-sphere integrals of the 2-divergence of vector fields $\ro^b\ro_a C^a{}_{bc} \mathring{\tilde{q}}^c{}_d$ and  $\ro^b\ro^c C^a{}_{bc} \mathring{\tilde{q}}_{ad}$, tangential to the $r=r_\circ$ 2-spheres, vanish by Stokes' theorem.} 
Therefore the equality refers only to the integrals, not to the integrands. Eq. (\ref{ricci-exp1}) immediately implies that the ADM energy can be expressed in terms of the leading-order (i.e.,\, $\f{1}{r^3}$-) part of the radial-radial component of curvature $\R_{ab}$ of the physical 3-metric\,\cite{aaam3+1}:
\be \label{ricci-exp2} E_{\rm ADM} = -\f{1}{8\pi G} \,\lim_{\ro\to \infty}\,\oint_{r=r_\circ}\!\!\!\! r\,\, (\ro^a \ro^b\, \R_{ab})\, \rmd^2V\, .\ee
This is in fact the expression of the ADM energy used in the Carlotto-Schoen analysis \cite{carlotto-schoen,schoen}. 
 
In the second step, we wish to further recast the expression in terms of Weyl curvature of the 4-metric $g_{ab}$ obtained by evolving the initial data. For this, one uses the  fact that the electric part $\E_{ab}= C_{acbd}\, \tau^b \tau^d$ of the 4-d Weyl tensor on the 3-manifold $\Sigma$ can be expressed in terms of the initial data and the Ricci tensor $R_{ab}$ of the 4-metric $g_{ab}$ as follows:
\be \E_{ab} = \R_{ab} - k_{ac} k_{b}{}^c + k k_{ab} - \f{1}{2}\big(q_a{}^c q_b{}^d + q_{ab} q^{cd})\, (R_{cd} - \f{1}{6} R_{mn}g^{mn} g_{cd})\,.\ee
Our asymptotic conditions  on the stress-energy tensor of matter fields imply the 4-d Ricci tensor falls of as $O(\f{1}{r^4})$ and those on the extrinsic curvature imply that the terms quadratic in $k_{ab}$ also fall-off as $O(\f{1}{r^4})$. Therefore it immediately follows\, \cite{aaam3+1} that the ADM energy can also be written as:
\be \label{weyl-exp} E_{\rm ADM} = -\f{1}{8\pi G} \, \lim_{\ro\to \infty}\,\oint_{r=r_\circ}\!\!\!\! r\,\, (\ro^a \ro^b\, \E_{ab})\, \rmd^2V\, . \ee
There is a covariant framework to describe spatial infinity in the 4-dimensional setting without any 3+1 decomposition \,\cite{aa-ein}. In that framework the full ADM 4-momentum $\admP$ is naturally expressed in terms of the 4-d Weyl tensor.  Eq. (\ref{weyl-exp}) is the `time' component of that expression of $\admP$ in the rest frame chosen by $\tau^a$ which is orthogonal to our Cauchy surface $\Sigma$.\, 

Finally, note that our fall-off conditions imply that the Ricci-tensor $R_{ab}$ of the 4-d metric $g_{ab}$ falls off as $\f{1}{r^4}$. Therefore, we can also express the ADM energy in terms of the Riemann tensor as
\be \label{riem-exp} E_{\rm ADM} = -\f{1}{8\pi G} \,\lim_{\ro\to \infty}\, \oint_{r=r_\circ}\!\!\!\! r\,\, (\ro^a \tau^b \ro^c\, \tau^d R_{abcd})\, \rmd^2V\,. \ee
It is this expression that naturally suggests a thought experiment to measure $E_{\rm ADM}$ in any space-time that is asymptotically flat at spatial infinity, including those obtained by the Carlotto-Schoen construction.\vskip0.1cm

\subsection{The Thought Experiment}
\label{s2.3}

To motivate the thought experiment, let us make a small detour and consider isolated system in \emph{Newtonian gravity}. Now the matter density has compact spatial support and the Newtonian potential is given by $\Phi = - GM/r + O(1/r^{2})$ in the asymptotic region outside sources. In full general relativity, it is the tidal force  $\nabla_{a} \nabla_{b}  \Phi$ that has a clean counterpart in terms of curvature. So, let us express mass $M$ in terms of the Newtonian tidal acceleration. For this, we can consider a large 2-sphere of radius $r$ surrounding the matter source, and a nearby concentric 2-sphere of radius $r-\epsilon$. Let us now consider a shell of (massive) test particles at rest on each of these two 2-spheres. Let us drop them at $t=0$. Then, to the leading order, the 2-spheres will continue to remain 2-spheres but their separation $\delta$ will increase because of tidal effects associated with the inhomogeneity of the field: particles on the inner 2-sphere will experience a slightly greater acceleration then those on the outer 2-sphere. To {leading order}, the increase in the separation is dictated by
\be \label{tidal} \ddot\epsilon\, = \, \big(\h{r}^{a}\, \h{r}^{b}\, \partial_{a} \partial_{b} \Phi\big)\, \epsilon \,=\, \f{2GM}{r^{3}} \, \epsilon  + O(\f{1}{r^4})\ee
where $\partial_{a}$ is just the 3-dimensional derivative operator of the Euclidean space. This equation leads to an expression of the Newtonian mass of the isolated system in terms of the tidal acceleration, as a limit of a 2-sphere integral
\be \label{newtonianM} M = \f{1}{8\pi G} \, \lim_{{r_{o}\to \infty}} \,\oint_{r=r_{o}}  r\, \h{r}^{a}\, \h{r}^{b}\, \partial_{a} \partial_{b} \Phi\,\, \rmd^{2}V  \ee
Our thought experiment to measure the central mass consists of  observing the increase in separation between radially separated  \,`partner particles'\, that lie on the two shells, and averaging it over angles. From Newtonian perspective the experiment --especially the ``averaging over angles" -- seems unnecessarily cumbersome. But to find the general relativistic analog of this thought experiment, averaging becomes essential as illustrated by the Carlotto-Schoen construction. 

We can now carry over this physical idea to general relativity by replacing the Newtonian tidal acceleration with the appropriate component of curvature that features in the geodesic deviation equation. Let us consider an isolated system in   
in general relativity represented by a space-time $(M, g_{ab})$  that is obtained by evolution of initial data on a 3-manifold $\Sigma$, satisfying the asymptotic conditions \smash{(i) - (iii)} listed above. On the 3-manifold $\Sigma$, introduce two 2-spheres  $S_1$ and $S_2$ that are metric 2-spheres w.r.t. $\qo_{ab}$ with radii $\rad_\circ$ and $\rad_\circ -\delta$, each endowed with a uniform distribution of (massive) particles, with 4-velocities aligned with $\tau^{a}$ \, the unit normal to $\Sigma$, at the initial time (represented by $\Sigma$). This is the general relativistic analog of dropping the particles from rest. Denote by $\epsilon(\theta,\varphi)$ the proper distance between points on $S_1$ and $S_2$ that are related by geodesics normal to $S_1$  (defined by $q_{ab}$). Let them fall freely, i.e., follow geodesic orbits w.r.t. the physical 4-metric $g_{ab}$.  By the geodesic deviation equation, the radial component of the tidal acceleration of these test particles will dictate the increase in separation between them.  At the initial time, we have: 
\be \h{r}_{a} \,(\ddot\epsilon\, \h{r}^{a}) = -\, \epsilon\,\, \h{r}^{a}\, \h{r}^{c} (\h{t}^{b}\h{t}^{d}\, R_{abcd}).\ee  
Therefore our discussion of Newtonian gravity motivates us to \emph{operationally} define the total energy of the system w.r.t. the rest frame defined by $\tau^a$ as:
\be \label{E-operational}  E_{\rm oper}\, =\, -\f{1}{8\pi G} \,\lim_{\rad_\circ \to \infty}\,\oint_{r= {\rad}\circ}\!\!\!\! r\,\, (\ro^a \tau^b \ro^c\, \tau^d R_{abcd})\, \rmd^2V\, \ee
But the right side is precisely the ADM energy  $E_{\rm ADM}$ of Eq. (\ref{riem-exp})! Thus, a measurement of the tidal acceleration at the instant defined by $\Sigma$ provides the operational meaning of the ADM energy of the initial data on $\Sigma$. Note that the thought experiment involves only tidal acceleration which is an intrinsic and essential feature of general relativity that holds for all gravitational fields, including the exotic ones in which space-time is flat outside an angular cone extending to infinity.  However, the \emph{integrand} in (\ref{E-operational}) can be locally negative; this is why it is possible to\, `screen'\, gravity as exemplified by the gluing techniques. Finally, while we used the fiducial flat metric $\qo_{ab}$ in the intermediate stage, its specific choice washes out in the limit $\rad_\circ \to \infty$. And, as we discussed in section \ref{s1}, we must take this limit in general relativity because the gravitational field itself contributes to the total energy. Thus, before taking the limit one obtains only an approximate value of $E_{\rm ADM}$: the thought experiment involves making a series of measurements using 2-spheres of increasing area and then taking the limit. Thanks to the positive energy theorem \cite{sy,ew},\, we are guaranteed that the total energy $E_{\rm ADM}$ is always positive so long as sources satisfy the energy condition specified above. 

To summarize, while the operational meaning of the original expression (\ref{ADMenergy}) of the ADM energy $E_{\rm ADM}$ is quite obscure, when recast in terms of the 4-d Riemann tensor one immediately sees that it can be measured by observing tidal acceleration of nearby spherical distributions of test particles in the asymptotic region. It is essential to perform the experiment in the asymptotic region --not just outside the matter source, as in Newtonian gravity-- because in general relativity, gravitational field itself contributes to the total energy (and hence to the total 4-momentum). A second contrast to Newtonian gravity is that one needs a whole 2-sphere worth of test particles surrounding the source because, in general relativity, the space-time  metric can be flat in large regions near spatial infinity even when $E_{\rm ADM}$ is strictly positive.  To extract its value, one needs an integration over all angles. By changing the asymptotic rest frame used in the thought experiment, one can measure different components of $\admP$ and thereby reconstruct the full 4-vector.

\section{The Bondi-Sachs Energy}
\label{s3}

Let us now turn to null infinity. As in section \ref{s2}, the thought experiment will be performed in the asymptotic region of the \emph{physical} space-time. Therefore, it is convenient to work with the Bondi-Sachs framework rather than Penrose's conformal completion. Thus, we will consider solutions that are asymptotically flat at null infinity in the sense that the physical metric $g_{ab}$ satisfies vacuum equations in the asymptotic region and admits the standard  Bondi-Sachs expansion there. \,\cite{bondi,sachs}.\,  In the absence of gravitational radiation at $\scrip$, the Bondi-Sachs energy is also expressible as an integral over the leading order (i.e. ``$1/r^3$") term in the electric part of the asymptotic Weyl curvature. Therefore, one can repeat the thought experiment of section \ref{s2}. The analysis has to be slightly modified to account for the fact that in the limit ${\ro\to \infty},$\, the 2-sphere approaches a cross-section of null infinity; the limit is taken along  a constant retarded time hypersurface $u = u_\circ$ rather than along a Cauchy surface $\Sigma$. Apart from the resulting modifications in the details, we can use the same thought experiment as in section \ref{s2.3}, as discussed in \,\cite{aasb}.

Tidal acceleration experienced by massive particles refers to the Coulombic aspect of the gravitational field (encoded in the curvature components that fall off as $1/r^3$). In presence of gravitational waves, the Bondi-Sachs energy has an additional term containing the radiative information encoded in metric coefficients that fall off as ${1}/{r}$.
Therefore a measurement of tidal acceleration of rings of  massive particles fails to  capture the full content of the Bondi-Sachs energy. However, it is possible to use instead properties of null rays  which can be thought of as trajectories of massless particles. Recall that the original expression of the ADM energy was in terms of the 3-metric and we had to recast it in terms of curvature that captures tidal acceleration directly. Similarly, the original Bondi-Sachs expression in terms of metric coefficients has to be recast to one that is more directly adapted to properties of null geodesics in the asymptotic region. As in the ADM case one proceeds in steps. The  first step of this procedure was already taken by Newman and Penrose, using null tetrads\,\cite{etnrp}.\, We will begin with that expression and further recast it using expansions of null normals to a suitable family of 2-spheres. The thought experiment to measure the Bondi-Sachs energy will involve the expansion of ingoing null rays.

\subsection{Preliminaries}
\label{s3.1}

Following Bondi and Sachs, let us introduce a null foliation\, $u= {\rm const}$\,\, in the asymptotic region of the given physical space-time $(M, g_{ab})$, consisting of outgoing null surfaces $\No$, labelled by $u=u_\circ$. Foliate each $\No$ by a family of 2-spheres given by $r= {\rm const}$, where $r$ is the area radius of the 2-sphere. Each of these 2-spheres has two future directed null normals. The outgoing null normal $\ell^a$ is tangential to $\No$ and normalized via $\Lie_\ell r =1$,\, and the ingoing one $n^a$, by the condition $g_{ab} \ell^a n^b =-1$. One can also carry out a conformal completion of $(M,g_{ab})$ \`a la Penrose\, \cite{rp}\, using $\Omega= \f{1}{r}$ as the conformal factor, so that the metric $\hat{g}_{ab} = \Omega^2 g_{ab}$ is well-defined at $\scrip$. In this conformal completion, the induced degenerate metric $\qo_{ab}$ at $\scrip$ is a round 2-sphere metric of unit radius;\, we have a ``Bondi conformal frame" at $\scrip$. 
For our thought experiment the conformal completion is not necessary. Nonetheless, as is often done in the literature,  it is convenient to refer to structures on $\scrip$ that \emph{would have resulted} had we carried out this completion. As one approaches $\scrip$, the chart $(u,r,\theta,\phi)$ breaks down in the limit, since $r \to \infty$. For limits of scalars this breakdown is harmless. But for limits of tensor fields, one cannot directly use the limiting behavior of their components in the $(u,r,\theta,\varphi)$ chart to draw reliable conclusions about the limits of the tensor fields themselves to $\scrip$.
\footnote{At spatial infinity, this precaution was unnecessary because one works with Cartesian charts of the flat 3-metric $\qo_{ab}$ to which the physical metric $q_{ab}$ approaches in the specified sense. But Cartesian coordinates are cumbersome to use in the approach to $\scrip$ along null hypersurfaces $u\!=\!u_\circ$.}\,\,
Therefore, to take these limits, we will use the chart $(u, \Omega, \theta, \varphi)$ and  take the\,  $\Omega \to 0$\, limits of components of various fields in this chart. Note, however, that we will work in the physical space-time $(M, g_{ab})$ (on which $\Omega$ is strictly positive) and all our fields will refer to the physical metric $g_{ab}$. But we will permit ourselves to say that in the limit $\Omega \to 0$, the family of 2-spheres $\Omega= \Omega_\circ$ on any one $\No$ defines a cross-section $S_\circ$ of $\scrip$, with $\h{\no}^a = \lim_{\Omega\to 0} n^a$ as its null normal within $\scrip$.\, In fact the limit $\h{\no}^a$ is  a Bondi-Metzner-Sachs (BMS) vector field on $\scrip$, representing a unit time translation. We will refer to \emph{any} smooth vector field on $M$, that would have converged to $\h\no^a$ at $\scrip$ in the conformal completion, as an \emph{asymptotic time-translation}. The Bondi-Sachs energy refers to this time-translation.

With this terminology at hand, the Newman-Penrose expression of Bondi-Sachs energy can be written as:
\be \label{BSE1} 
E_{\rm BS}[S_\circ] = - \f{1}{8\pi G}\, \lim_{\Omega_\circ \to 0}\,\oint_{\Omega=\Omega_\circ} \!\!\!\Omega^{-1}\, 
\big[C_{abcd}\, n^a \ell^b n^c \ell^d\,+\, \Omega^{-1}\,\sigma^{(\ell)}_{ab}\, (\Lie_{n} \sigma_{cd}^{(\ell)})\,\, \tl{q}^{ac}\, \tl{q}^{bd}\big]\,\, \rmd^2 V\, .
 \ee
Here $\tilde{q}_{ab}$ is the intrinsic metric on the 2-spheres $\Omega=\Omega_\circ$ \, on the null sheet $\No$\, and {\smash{$\sigma_{ab} =  {\rm TF}\,\, \tl{q}_a{}^c \tl{q}_b{}^d\,\nabla_c \ell_d$}} $\equiv (\tl{q}_a{}^c \tl{q}_b{}^d  - \f{1}{2} \tl{q}^{ab}\, \tl{q}^{cd})\,\nabla_c \ell_d$  is the shear of the outgoing null normal $\ell^a$ to the $\Omega\!=\!\Omega_\circ$\, 2-spheres, where TF stands for ``trace-free part of". In spite of the explicit factors of $\Omega^{-1}$ and an additional implicit factor of $\Omega^{-2}$ in $\rmd^2 V$, the limit is well defined because the Bondi-Sachs boundary conditions and vacuum field equations near $\scrip$ imply that the term in square brackets is $O(\Omega^3)$. The limit to $\scrip$ of\, $2 \Lie_{n} \sigma_{ab}^{(\ell)}$\, coincides with the News tensor $N_{ab}$\,\cite{aass2}.

Just as the expression of the ADM energy was first given in terms of  limits of components of the 3-metric as one approaches spatial infinity \cite{adm}, the Bondi-Sachs energy was first expressed using limits of components of the 4-metric at null infinity \cite{bondi,sachs}. The Newman-Penrose recasting (\ref{BSE1}) using Weyl curvature and geometric properties of null congruences is very similar to the recasting (\ref{weyl-exp}) of the ADM expression in terms of the Weyl tensor. Indeed, the first term in (\ref{BSE2}) is the same as that in (\ref{weyl-exp}): In the Newman-Penrose framework, it is the component $2{\rm Re} \Psi_2$ of the Weyl curvature in both cases, albeit in the ADM case the limit is taken as the 2-sphere expands out to infinite radius along a Cauchy surface, while for the Bondi-Sachs energy the 2-sphere approaches a cross-section of $\scrip$. The second term in (\ref{BSE1}), however, does not have an analog in the ADM case. It vanishes in absence of radiation --i.e. if the news $N_{ab}$ vanishes-- and then, as we already commented, one can repeat the thought experiment of section \ref{s2}\, \cite{aasb}.\,  But in presence of gravitational radiation, that thought experiment \emph{does not} provide a measurement of Bondi energy. In section \ref{s3.2} we will further recast (\ref{BSE1}) using only the expansions of null rays and use that expression in section \ref{s3.3} to introduce a thought experiment that faithfully captures the effect of Bondi energy, including the radiative term.

\subsection{Recasting the Bondi-Sachs Energy in terms of Null Expansions}
\label{s3.2}

We need to recast the expression (\ref{BSE1}) to a form that is well-adapted to a thought experiment. As in section \ref{s2}, we will proceed in three steps.

Let us begin by noting that the boundary conditions and algebraic Bianchi identities imply that $N_{ab}$ is also the limit to $\scrip$ of $ -2 \Omega \sigma^{(n)}_{ab}$ where $\sigma^{(n)}_{ab} = {\rm TF}\,\, \tl{q}_a{}^c \tl{q}_b{}^d \,\nabla_c n_d$ is the shear of the \emph{ingoing} null normals. Therefore, the right side of (\ref{BSE1}) can also be written as:
\be \label{BSE2} E_{\rm BS} [S_\circ] = - \f{1}{8\pi G}\, \lim_{\Omega_\circ \to 0}\,\oint_{\Omega= \Omega_\circ}\!\!\!\Omega^{-1}\, \big[C_{abcd}\, n^a \ell^b n^c \ell^d\,-\,\sigma^{(\ell)}_{ab} \sigma_{cd}^{(n)} \,\, \tl{q}^{ac}\,\tl{q}^{bd}\big]\, \rmd^2 V\, . \ee
In the next step, we will use a 2+2 decomposition of the 4-dimensional curvature. This is analogous to the more familiar 3+1 decomposition in which the Gauss-Codazzi equations relate the intrinsic and extrinsic curvature of a 3-d submanifold $\Sigma$ to projections of the 4-d curvature tensor. Now, any 2-sphere $S$ in a 4-d space-time is naturally equipped with two null normals $\ell^a$ and $n^a$ (which can always be required to satisfy the normalization condition $\ell^a n_a = -1$). Using these normals one can carry out a 2+2 decomposition and relate projections of the 4-d curvature tensor to the intrinsic curvature of the metric on $S$ and the extrinsic curvatures of the two null normals, each of which admits a natural  decomposition into trace terms (the expansions $\theta_{(\ell)},\, \theta_{(n)}$)\, and trace-free terms\, (the shears $\sigma_{ab}^{(\ell)},\, \sigma_{ab}^{(n)}$). If the 4-metric $g_{ab}$ satisfies the vacuum Einstein's equation, the decomposition yields a constraint: 
\be \label{GC}
C_{abcd}\,n^a \ell^b n^c \ell^d\, - \,  \sigma_{ab}^{(\ell)}\, \sigma^{ab}_{(n)}\,+\,{\textstyle{\f{1}{2}}}\, \big(\, {}^{\scriptscriptstyle{2}}\tl{R} \,+\,\,\theta_{(\ell)}  \theta_{(n)} \big) \,=\, 0 , \ee
where ${}^{\scriptscriptstyle 2}\tl{R}$ is the scalar curvature of the intrinsic 2-metric $\tl{q}_{ab}$ on $S$. This is an exact identity that holds outside sources; no asymptotic conditions have been used. Since the first two terms on the left side constitute 
the integrand of the Bondi-Sachs energy (\ref{BSE2}), one can express $E_{BS} [S_\circ]$ also as an integral of the scalar curvature and the two null expansions:
\be \label{BSE3} E_{\rm BS} [S_\circ] =  \f{1}{16\pi G}\, \lim_{\Omega_\circ \to 0}\,\oint_{\Omega= \Omega_\circ}\!\!\!\Omega^{-1}\, \big[ {}^{\scriptscriptstyle 2}\tl{R} \,+\,\,\theta_{(n)}  \theta_{(\ell)} \big]\,\, \rmd^2 V\, . \ee
The first term $ {}^{\scriptscriptstyle 2}\tl{R}$ on the right side can be readily integrated using the Gauss-Bonnet formula. However, since the \emph{integrand} in the second term involves products of null expansions, the expression is still not well-suited for a thought experiment (see Remark 2 at the end of the section \ref{s3.3}). Therefore, we have to further recast (\ref{BSE3}) to a more convenient form using the Bondi-Sachs boundary conditions and the vacuum Einstein's equations.

Let us begin by spelling out the implications of Bondi-Sachs boundary conditions that are needed in this last step: Using the chart $(u,\Omega,\theta, \varphi)$, the Bondi-Sachs boundary conditions imply
\ba 
\label{BSfalloff} & g_{uu} = -1 +O(\Omega); \qquad & g_{u\Omega} = \Omega^{-2}+O(1); \nonumber\\
& g_{uA} = O(1); & g_{AB} = \Omega^{-2} \qo_{AB} + \Omega^{-1} C_{AB} +O(1)\, , \ea
where $A,B =1,2$ stand for $\theta, \varphi$,\, and $\qo_{AB}$ is the unit 2-sphere metric. The Bondi-Sachs fall-off conditions on the metric coefficients also imply that coordinate derivatives of these metric  components satisfy the fall-off conditions compatible with Eq. (\ref{BSfalloff})  (e.g., $\partial_\Omega\, g_{uu} = O(1) + O(\Omega)$, \,\, $\partial_u g_{uu} = O(\Omega),\,\, \partial_A g_{uu} = O(\Omega)$).  These fall-offs imply that the two expansions have the following behavior
\footnote{\label{fn} The relative factor of $2$ in the leading terms comes from the conventions: $\ell^a$ is normalized so that $\ell^a\,\partial_a\, r =1$ and $n^a$ is then normalized by demanding $\ell^a n_a =\,-1$.}
\be \label{expansions} \theta_{(n)} = - \Omega + O(\Omega^2) \qquad {\rm and} \qquad  \theta_{(\ell)} = 2\Omega + O(\Omega^3). \ee
The absence of a term $O(\Omega^2)$ in the fall-off of $\theta_{(\ell)}$ is a consequence of the absence of\, terms of the order $O(\Omega^{-1})$\, in  $g_{u\Omega}$,\, which in turn is a consequence of Einstein's equations. Next, we note that since $\oint_S {}^{\scriptscriptstyle{2}}\tl{R}\, \rmd^2V$ is the Gauss invariant that equals $8\pi$ for any 2-sphere $S$, and since\, $\Omega^{-1}\! = r$,\, the area radius, we have
\be \label{gauss} 
\Omega^{-1} \oint_S \big( {}^{\scriptscriptstyle 2}\tl{R} \,-\, 2\Omega^2\big)\, \rmd^2V = 0 
\ee
for any $S$ in the Bondi-Sachs family of 2-spheres on the $u=u_\circ$ null surface, even before taking the limit in (\ref{BSE3}). Using (\ref{gauss}) in \eqref{BSE3}, we have:
\be E_{\rm BS} [S_\circ] = \f{1}{16\pi G}\, \lim_{\Omega_\circ \to 0}\,\oint_{\Omega= \Omega_\circ}\!\!\! \big[ 2 \Omega\, +\, \Omega^{-1}\, \theta_{(n)} \theta_{(\ell)}) \big]\, \rmd^2 V\, . \ee
By adding and subtracting $2\theta_{(n)}$ in the integrand and rearranging the terms, we obtain
\ba \label{BSE4} E_{\rm BS} [S_\circ] &=& \f{1}{16\pi G}\, \lim_{\Omega_\circ \to 0}\,\oint_{\Omega= \Omega_\circ}\!\!\! \big[2(\theta_{(n)} + \Omega)\, +\, 
      \Omega^{-1} \theta_{(n)} (-2\Omega \,+\, \theta_{(\ell)})\,\big] \, \rmd^2 V \nonumber\\
&=& \f{1}{8\pi G}\, \lim_{\Omega_\circ \to 0}\,\oint_{\Omega= \Omega_\circ}\!\!\! (\theta_{(n)} + \Omega)\, \rmd^2 V\, , \ea
where in the last step we used the fact that the second term in the integrand of the first equation vanishes in the limit 
because of the fall-off conditions in (\ref{expansions}): the term\,{\smash{ $\theta_{(n)}\,(-2\Omega \,+\, \theta_{(\ell)})$}}\,\, is\,\, $O(\Omega^4)$,\, and the limit of $\Omega^{-2} \rmd^2V$ is well-defined (and equals the volume element of the unit 2-sphere). 
The final expression shows that it is possible to determine the Bondi-Sachs energy $E_{\rm BS} [S_\circ]$ using a single geometrical field: the expansion $\theta_{(n)}$ of the inward pointing null normal $n^a$. This will pave the way for the thought experiment to measure $E_{\rm BS} [S_\circ]$.

We will conclude by providing physical intuition as to why the angular average of $(\theta_{(n)} + \Omega)$ can be thought of as total energy of the isolated system. Recall that coordinates $(u, \Omega, \theta, \varphi)$ are chosen using properties of the physical metric $g_{ab}$. In particular,  $u\!=\!u_\circ$ is a null hypersurface $\No$, and $r_\circ\!=\! \Omega_\circ^{-1}$ is the area-radius of 2-spheres $S$ given by $u\!=\!u_\circ, r\!=\!r_\circ$, both w.r.t. $g_{ab}$. Now, using the limiting values of components of $g_{ab}$ in these coordinates, one can obtain a Minkowski metric $\go_{ab}$ in the asymptotic region of the spacetime to which $g_{ab}$ approaches as $\Omega\to 0$. One can easily check
that $\No$ is  a null surface and $S$ is 2-sphere with area radius $r_\circ$,\, \emph{also} w.r.t. the Minkowski metric $\go_{ab}$. The two metrics $g_{ab}$ and $\go_{ab}$ share the vector field $\ell^a$ but the ingoing normal $\no^a$ defined by $\go_{ab}$ differs from $n^a$. In particular, the expansion $\mathring\theta_{(\no)}$\, of\,  $\no^a$ w.r.t. $\go_{ab}$ is just $-\Omega$. Therefore the integrand in\, (\ref{BSE4})\, can also be understood as the difference between the physical expansion and the corresponding expansion in the flat background,
\be \label{BSE4bis} 
E_{\rm BS} [S_\circ] =\f{1}{8\pi G}\, \lim_{\Omega_\circ \to 0}\,\oint_{\Omega= \Omega_\circ}\!\!\! (\theta_{(n)} - \mathring\theta_{(\no)})\, \rmd^2 V. 
\ee

%
The Bondi-Sachs energy $E_{\rm BS}[S_\circ]$ is given by the angular average of the `extra' null expansion caused by the curvature in $g_{ab}$  \emph{over and above} that in flat space, in the limit $\Omega \to 0$. It is this physical manifestation of the total energy in the isolated system that is captured by expression (\ref{BSE4}). Note also that the 2-sphere integral of $\theta_{(n)}$ by itself diverges in the limit: the `flat spacetime subtraction' makes it finite. Remarkably, this subtraction is automatically supplied by the codimension-2 Gauss-Codazzi equation \eqref{GC}.
%
\\

\emph{Remarks}:\\ 
\indent (i) Recall that the 2-spheres $\Omega=\Omega_\circ$ in (\ref{BSE3}) and (\ref{BSE4}) lie on the $u=u_\circ$ null hypersurface $\mathcal{N}_\circ$. While $\ell^a$ is tangential to $\mathcal{N}_\circ$, $n^a$ is transverse to it. Although $\theta_{(n)}$ involves derivative of $n^a$, it is sufficient to specify $n^a$ just on $\mathcal{N}_\circ$: thanks to the  projector $\tl{q}^{ab}$\,  --the metric on the 2-spheres-- \, in the definition $\theta_{(n)} = \tl{q}^{ab} \nabla_a n_b$ of the expansion,\, $\theta_{(n)}$ is insensitive to the way in which $n^a$ is extended away from $\mathcal{N}_\circ$.

(ii) Since $\theta_{(n)} <0$, the \emph{convergence} $|\theta_{(n)} (r_\circ)|$  of ingoing null rays decreases as $E_{\rm BS}$ increases, at fixed $r_\circ$ in the asymptotic region. For example, in the Schwarzschild solution representing a star, the convergence is given by $|\theta_{(n)}(r_\circ)| = \f{1}{r_\circ} (1 - \f{2GM}{r_\circ})$, which decreases as $M$ increases. This may seem counter-intuitive because one would expect that a larger mass would result in a larger convergence of inward pointing null rays. Note, however, that the larger the mass, the larger is the proper spatial separation between the surface of the star and the surface $r=r_\circ$. Therefore, comparing values of $|\theta_{(n)}|$ at fixed area-radius $r$ can be misleading. A more appropriate way to compare the convergence of rays in different spacetimes is to use the Kretschmann scalar as a benchmark, namely compare
$|\theta_{(n)}(r_1)|$  in the asymptotic region of the first space-time, and $|\theta_{(n)}(r_2)|$ in the second, such that the space-time curvature on the 2-sphere $r=r_1$ in the first space-time and $r=r_2$ in the second is the same. In the  Schwarzschild example one evaluates $|\theta_{(n)}|$ with mass $M_1$ on a 2-sphere $r=r_1$ and with mass $M_2$ on a 2-sphere $r=r_2$ such that $M_1/r_1^3 = M_2/r_2^3$. Then, one does find that the convergence increases with the Schwarzschild mass, just as one would intuitively expect. 

\subsection{The Thought Experiment}
\label{s3.3}

Eq.(\ref{BSE4}) is the null infinity analog of Eq. (\ref{riem-exp}) that was tailored to spatial infinity. Just as the thought experiment  for the ADM energy is based on (\ref{riem-exp}), the one to measure the Bondi-Sachs energy will be based on (\ref{BSE4}). That is, the desired experiment will provide a measurement of the average of $(\theta_{(n)} + \Omega)$ over angles, in the limit $\Omega \to 0$.

To explain the idea behind the thought experiment, let us first set it up in Minkowski space $(M, \go_{ab})$. Although this case is physically trivial since $E_{\rm BS}$ vanishes, the exercise provides a convenient starting point to set the stage. Consider a large, static, round material sphere $\S$ of radius $r_\circ$ in the asymptotic region of $(M, \go_{ab})$ (As in section \ref{s2},\, ${\S}$ is chosen to be a test body whose effect on the gravitational field is negligible.) Its 4-velocity vector field $\tau^a$ will be the restriction to $\S$ of a unit time-translation Killing field $t^a$ adapted to the rest frame of $\S$. Let us illuminate $\S$ 
at time $t=t_\circ$, where $t$ is the time measured in the rest frame of $\S$. There will be two light fronts, an outgoing one which expands and the ingoing one that contracts.\, The null-sheet spanned by the expanding light front, $\No$, will intersect $\scrip$ at some cross-section $S_\circ$ (which, incidentally, will be shear-free since $\S$ is round). Our thought experiment to determine $E_{\rm BS} [S_\circ]$ involves two measured quantities. The first is the area $A_{\S} = 4\pi r_\circ^2$ of the material 2-sphere $\S$, which is also the area $A (t_\circ)$ of both light fronts at time $t_\circ$. The second will result from a set of of measurements of the area $A(t_\circ+\epsilon)$ of the \emph{ingoing} light front at instants $t_o +\epsilon$ in the rest frame of ${\S}$. That area will be $4\pi (r_\circ- \epsilon)^2$ so the rate of change in area w.r.t. $t$ will be $\dot{A}(t_\circ) =  -\f{2}{r_\circ} A(t_\circ)\, \equiv\, - 2\Omega_\circ\, A(t_\circ)$. This is our second measured quantity which an be obtained, for instance, by a distribution of light sensors in a small neighborhood of ${\S}$, inside it. 

\begin{figure}
\centering 
\includegraphics[width=6.5cm]{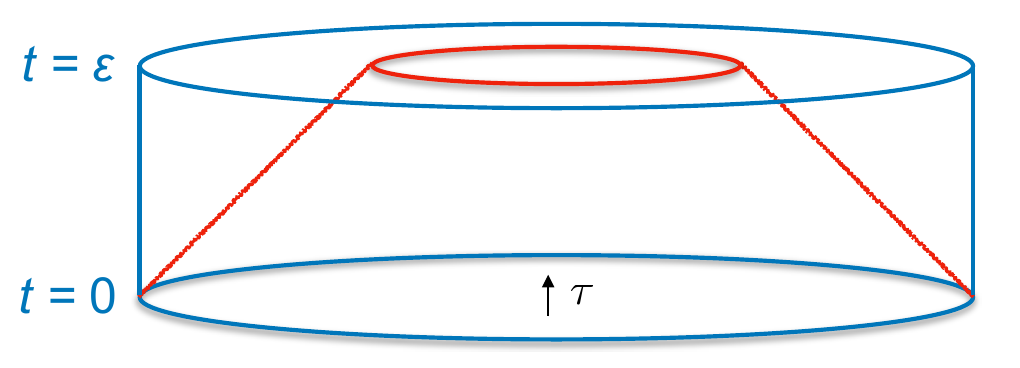} \hspace{1cm}\includegraphics[width=7.5cm]{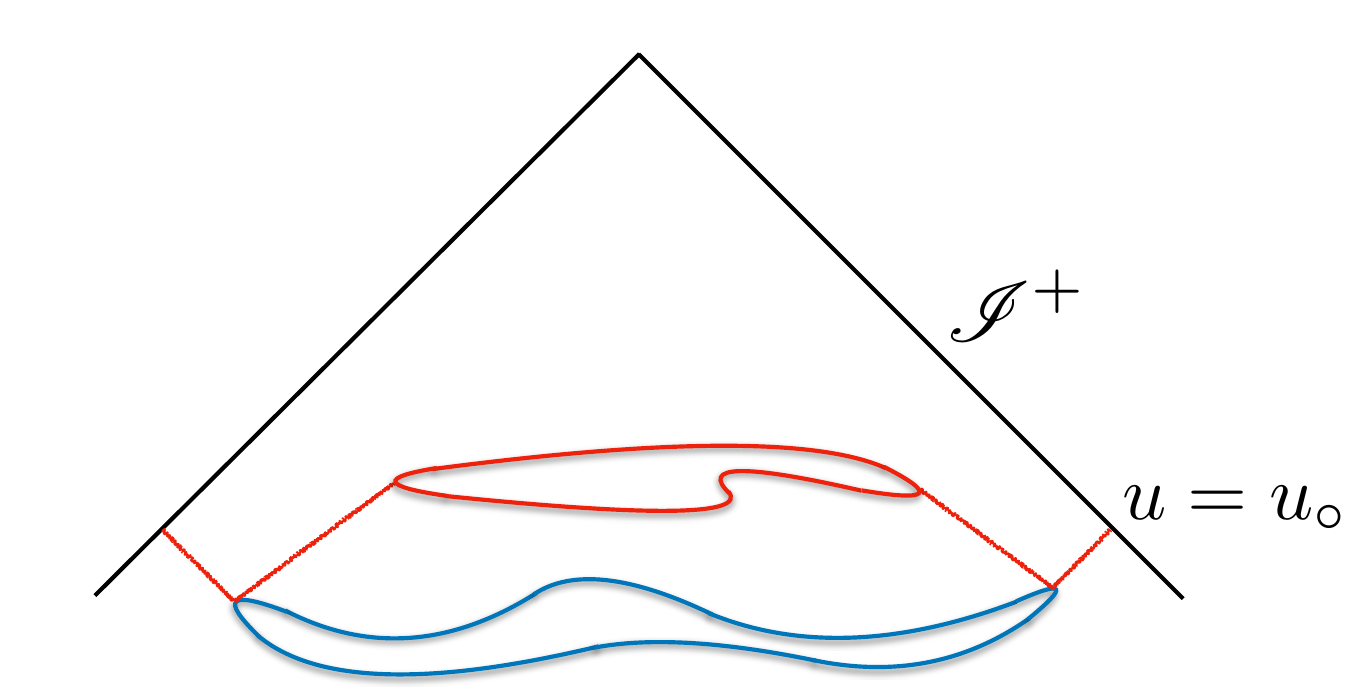}
\caption{\footnotesize{{\emph{Left Panel:} Minkowski space. At time $t\!=\!0$, we illuminate a static 2-sphere $\S$ with a large radius $r_\circ$. The figure shows the world tube of the 2-sphere (in blue) and the light front tangential to the \emph{ingoing} null normal $n^a$ to $\S$. At $t\!=\!0$, the area of this light front is $4\pi r_\circ^2$ and at $t\!=\! \epsilon$ it would shrink to $4\pi (r_\circ-\epsilon)^2$.\, These measurements of areas of the ingoing light fronts provide the average (over angles) of the expansion $\mathring\theta_{(n)}$ at time $t\!=\!0$.\\ 
\emph{Right Panel:} A generic asymptotically flat space-time. Again the 2-sphere $S$ in the asymptotic region of space-time is illuminated at an initial instant. The outgoing light-front, representing a retarded instant of time $u=u_\circ$, intersects $\scrip$ at a cross-section $S_\circ$. At instants $\epsilon$ later we measure the areas of the ingoing light front. These two measurements provide us the observed value of the average of $\theta_{(n)}({r_\circ})$ over angles. By repeating the experiment  for a family of 2-spheres on the outgoing light front $\No$, in the limit we obtain the value of the Bondi-Sachs energy associated with the cross-section $S_\circ$ of $\scrip$ via Eq.(\ref{BSE4}).
}}}
\label{Fig1}
\end{figure}

To relate the observed quantities $A$ and $\dot{A}$ to the mathematical quantity $\theta_{(n)}$ that determines $E_{\rm BS} [S_\circ]$ via (\ref{BSE4}), we just have to use the definition of the expansion $\theta_{(n)}$.  Since $\S$ is round, we have:  $\theta_{(n)} (t_\circ) = \f{\dot{A}}{2A}(t_\circ)$. 
\footnote{The factor of 2 arises because  in Minkowski space the Bondi-Sachs normalization conventions  discussed in footnote \ref{fn} imply $n^a = \f{1}{2} (t^a - \hat{r}^a)$, where $t^a$ is a unit time translation and $\h{r}^a$ unit radial vector in the orthogonal spatial slice.}
Hence the Bondi-Sachs energy (\ref{BSE4}) can be expressed in terms of \emph{observed quantities} as 
\be \label{BSE5} E_{\rm BS} [S_\circ] =  \f{1}{8\pi G}\, \lim_{\Omega_\circ \to 0}\,\oint_{\Omega= \Omega_\circ}\! \Big(\f{\dot{A}}{2A} + \Omega\Big)\,  \rmd^2 V\, .
\ee 
As expected, $E_{\rm BS} [S_\circ]$ vanishes because for a round 2-sphere of radius $r_\circ$ in Minkowski space, $\theta_{n}(t_\circ)  = - 1/r_\circ$, whence $\dot{A}(t_\circ)/2A(t_\circ) = -\Omega_\circ$. In fact this implies that the right side of (\ref{BSE5}) vanishes already at $\Omega=\Omega_\circ$, even before taking the limit. This simplification occurs because we chose $\S$ to be a round sphere. With an arbitrary 2-sphere, the integral vanishes only in the limit. 

We can now consider general space-times $(M, g_{ab})$ satisfying the Bondi-Sachs asymptotic conditions. Each of these space-times admits a chart $(u,\Omega, \theta, \varphi)$ in the asymptotic region in which the metric components satisfy (\ref{BSfalloff}). Choose a large, material 2-sphere $\S$ of area-radius $r_\circ$, illuminate it instantaneously at $t=t_\circ$ and, as before, denote by $\No$ the outgoing light sheet that intersects $\scrip$ at a cross-section  $S_\circ$. Now $\S$ will generically not be round and $S_\circ$ will not be shear-free.

The Bondi-Sachs ansatz provides a foliation of $\No$ by 2-spheres (labeled by their area radius $r$) which is preserved by the 1-parameter family of diffeomorphisms generated by the null normal $\ell^a$ to $\No$. It also equips each of these 2-spheres with a future directed, ingoing null normal $n^a$ satisfying $n^a \ell^b g_{ab} = -1$. Using these vector fields let us introduce a unit time-like null normal $\tau^a$ to the family of 2-spheres via $\tau^a = \ell^a/2 +n^a$, and a space-like unit normal $\rho^a = \ell^a/2 - n^a$ which is also orthogonal to $\tau^a$.
Since $\ell^a \partial_a = \partial_r = - \Omega^2 \partial_\Omega$,\,\, $\ell^a$ vanishes in the limit to $\scrip$ along $\No$. Hence we have $\lim_{\Omega\to 0} (\tau^a - n^a) =0$. Now, the limit to $\scrip$ of $n^a$ yields the restriction to $S_\circ$ of a Bondi-Sachs time-translation. Thus, the vector field $\tau^a$  we defined is the restriction to $\No$ of an asymptotic BMS time-translation. The 2-sphere $\S$ follows the integral curves of this asymptotically static Killing field; it is as\, `static'\, as it can be in asymptotic region of the given curved space-time.

As in the case of Minkowski space, the idea is to measure areas of the contracting light fronts and compute ${\dot{A}}/{2A}$ at time $t_\circ$. (Here and in what follows\, `dot' \,refers to the directional derivative along $\tau^a$.) Since $n^a = \f{1}{2}(\tau^a - \rho^a)$, and the 4-velocity of $\S$ is given by $\tau^a$, in there rest frame of $\S$, the contracting light front moves inward along $- \rho^a$. Therefore, it follows from the definition of $\theta_{(n)}$ that  $\oint_{\S} \theta_{(n)} \rmd^2 V=  \f{1}{2}\,\dot{A}$. Consequently from the first equation in (\ref{BSE4}) we have:
\be \label{BSE6} E_{\rm BS} [S_\circ] = \f{1}{8\pi G}\, \lim_{\Omega_\circ \to 0}\,\oint_{\Omega= \Omega_\circ}\! \Big[(\f{\dot{A}}{2A} + \Omega)\, +\,\Omega^{-1} \theta_{(n)} (-2\Omega \,+\, \theta_{(\ell)})\,\Big] \, \rmd^2 V\,.\ee
The Bondi-Sachs fall-off conditions (\ref{expansions}) on $\theta_{(n)}$ and $\theta_{(\ell)}$ imply that 
{\smash{$\theta_{(n)} (-2\Omega \,+\, \theta_{(\ell)})$} is $O(\Omega^4)$ but, in contrast to the situation in Minkowski space for round 2-spheres, generically it is non-zero for $\Omega = \Omega_\circ \not=0$. Therefore the integral in (\ref{BSE6}) only provides an approximate value of $E_{\rm BS}$ before taking the limit. To obtain the exact value we need to use a limiting  procedure by introducing a sequence of material 2-spheres $\S(r_n)$ of larger and larger areas that lie in the $r={\rm const.}$ foliation of the null sheet $\No$, with 4-velocities aligned with the asymptotic time-translation defined by $\tau^a$. In the limit $\Omega\to 0$, the integral in (\ref{BSE6}) provides the Bondi-Sachs energy $E_{\rm BS}[S_\circ]$. This operational method to measure the Bondi-Sachs energy relies on the fact that it is possible to express $E_{\rm BS} [\S_\circ]$ solely in terms of $\theta_{(n)}$, in the limit $\Omega \to 0$.} 
\bigskip

\emph{Remarks}:

1. Generically, the null rays emanated from a 2-sphere $S$ in the space-time interior encounter caustics. This is true even in the Schwarzschild space-time if $S$ is not chosen to be a round 2-sphere. However, in our analysis the material 2-spheres $\S$ always lie in the asymptotic region. Therefore, the expanding null geodesics will not encounter any caustics and the light-fronts $\N$ will be smooth. The contracting null rays \emph{will} generically encounter them once they leave the asymptotic region. But since we only need their expansion $\theta_{(n)}$ on $\N$, these caustics are not relevant for our analysis.

2. The integral in (\ref{BSE3}) for finite $\Omega_\circ$ (i.e., before taking the limit)  is sometimes referred to as Hawking's quasi-local mass because this expression appears in a parenthetical remark (after Eq. (8.1)) in \cite{swh}. However, it is more appropriate to refer to it as quasi-local energy, since  even in the limit $\Omega_\circ \to 0$, what it yields is the Bondi-Sachs \emph{energy} $E_{\rm BS}$ associated with the BMS time-translation defined by $n^a$ at $\scrip$, rather than the Bondi-Sachs \emph{mass}. One can choose Bondi coordinates $(u, r,\theta,\varphi)$ so that $\partial_u\!\!\mid_{\scrip}\, = n^a \partial_a\!\!\mid_{\scrip}$  is \emph{any} desired BMS time-translation at $\scrip$ and the result would yield the  corresponding Bondi energy which would vary with the choice of $n^a$; it would equal the Bondi-Sachs mass only if the Bondi-Sachs 3-momentum in the rest frame defined by $n^a\!\!\mid_{\scrip}$ vanishes. 
This was indeed the case in Hawking's paper \cite{swh}, whose goal was to extend the Bondi-Sachs framework to study gravitational waves on a dust-filled Friedmann-Lema\^itre  space-times (with negative spatial curvature). In this case, there \emph{is} a preferred rest frame and the total energy defined in that frame is the same as the total mass. In asymptotically flat space-times, on the other hand, we do not have a preferred rest frame and we need to distinguish mass and energy. Therefore, in the asymptotically flat context, the integral in (\ref{BSE3}) is Hawking's quasi-local energy. 

This definition of energy has the advantage that it does not require \emph{separate} normalizations of $\ell^a$ and $n^a$; one only needs $g_{ab}\,\ell^a n^b =-1$, and the rescaling freedom $\ell^a \to \lambda \ell^a;\,\, n^a \to \f{1}{\lambda} n^a$ is maintained. Could we have not used this\, `more invariant'\, expression in a thought experiment? The answer is in the negative: because its  \emph{integrand} contains a product $\theta_{(\ell)} \theta_{(n)}$,\, the integral is not  related to the change in area in any simple way. (If instead we had the product {\smash{$(\oint_{\Omega=\Omega_\circ} \theta_{(\ell)} \rmd^2 V) \times (\oint_{\Omega=\Omega_\circ} \theta_{(n)} \rmd^2 V)$} of \emph{integrals,}\, it would have had a simple operational meaning: it would have been the product of rates of area change in the two directions. Then, we could have used that expression for the thought experiment.) Also, for the same reason, the expression (\ref{BSE3}) cannot be directly used to construct a quantum operator representing the Bondi-Sachs energy. On the other hand, as we discuss in section \ref{s4}, one \emph{can} use (\ref{BSE4}) to promote (\ref{BSE4}) to an operator in loop quantum gravity.

3. The passage from (\ref{BSE3}) to (\ref{BSE4}) introduces an asymmetry between $\ell^a$ and $n^a$: The expression (\ref{BSE4}) of $E_{\rm BS}$ involves only $\theta_{(n)}$ and not $\theta_{(\ell)}$. Could we have perhaps obtained another expression that involves only $\theta_{(\ell)}$? Such an expression  would also be well-suited for a thought experiment (and in loop quantum gravity as well). Interestingly, this is not possible! Although the right side of (\ref{BSE3}) is symmetric in $\theta_{(n)}$ and $\theta_{(\ell)}$, in the passage to  Eq.(\ref{BSE4}) we made a crucial use of the fact that they have different asymptotic behaviors (see \eqref{expansions}). The fact that there is no term $O(\f{1}{r^2})$ in the asymptotic expansion of $\theta_{(\ell)}$ was used crucially to remove it in the final expression (\ref{BSE4}). And this asymmetry in the asymptotic behavior of the two expansions can in turn be traced back to the fact that, given the Bondi-Sachs expansion of the 4-metric, Einstein's equations imply that the potential term of the order  $O(\Omega)$ in $g_{u\Omega}$ vanishes, and we have $g_{u\Omega} = \Omega^{-2} + \Omega(1)$ (see Eq. \ref{BSfalloff}).

4. As we saw in Section \ref{s2}, tidal acceleration is a natural tool to provide an operational meaning to the ADM energy. However, as we saw, in presence of gravitational waves at $\scrip$ the tidal acceleration fails to capture the full content of the Bondi-Sachs energy $E_{\rm BS}$. We were led to use instead the expansion $\theta_{(n)}$ of inward pointing null rays. A natural question then is: Could we not use $\theta_{(n)}$ also to give an operational meaning to $E_{\rm ADM}$? If so, we would be able to use the same thought experiment to measure $E_{\rm ADM}$ and $E_{\rm BS}$;\,\, in the first case we would let the expanding 2-spheres $\S$ approach $i^\circ$ along a Cauchy surface, and in the second case, consider a family of 2-spheres that approaches a cross-section of $\scrip$.%
\footnote{Of course if one uses the fact \cite{aaam-prl} that the ADM 4-momentum is the past limit of the Bondi-Sachs 4-momentum in space-times that are asymptotically flat in both null and spatial directions, the answer is trivially in the affirmative. The question is whether we can base a thought experiment using $\theta_{(n)}$ to measure the ADM energy in space-times using asymptotic flatness \emph{only} in spatial directions.} 
Interestingly, as we explain below, the answer is in the negative! 

Since the basic identity (\ref{GC}) that led us to the expression (\ref{BSE4}) of $E_{\rm BS}$  in terms of $\theta_{(n)}$ holds for all 2-spheres, we can indeed use it for the family of 2-spheres with increasing radii that lie in the asymptotic region of a Cauchy surface. However, the situation at spatial infinity is different from that at null infinity in two aspects. First, since the integrand of $E_{\rm ADM}$ in (\ref{weyl-exp}) involves only a component of the weyl tensor which can be rewritten as $C_{abcd}n^a\ell^b n^c \ell^d$;\, it does not contain the shear terms that are present in the expression (\ref{BSE1}) of $E_{\rm BS}$. Therefore, the identity (\ref{GC}) implies that the integrand  $r\,r_\circ^ar_\circ^b\,\E_{ab}$ in the expression (\ref{weyl-exp}) of the ADM energy can be rewritten as\,\, $\sigma_{ab}^{(\ell)}\, \sigma^{ab}_{(n)}\,-\,{\textstyle{\f{1}{2}}}\, \big(\, {}^{\scriptscriptstyle{2}}\tl{R} \,+\,\,\theta_{(\ell)}  \theta_{(n)} \big)$:\, in contrast to  the analogous integrand of the Bondi-Sachs energy, we now have products of the two \emph{shears}, in addition to the product of the two expansions. Also, now the fall-offs of the shear and expansion of $\ell^a$ turn out to be the same as their analogs for $n^a$. Therefore, even in the limit $r \to \infty$ we are left with \emph{both} shears and \emph{both} expansions in the expression of $E_{\rm ADM}$. As a consequence --in contrast to $E_{\rm BS}$--\,  $E_{\rm ADM}$ cannot be expressed purely in terms of $\theta_{(n)}$. This discussion brings out the subtle nature of the structure at spatial infinity $i^\circ$; it distinguishes limits obtained by approaching $i^\circ$ directly along space-like directions, and those obtained by first going to $\scrip$ and then approaching $i^\circ$ along null directions \cite{ak1}.

5. One may ask why we focused on thought experiments. It is because our primary purpose is to bring out the \emph{global} nature of the notion of `total energy of an isolated system'. These experiments  are well beyond the reach of today's technology precisely for that reason: they face squarely the fact that the expressions of energy require an integral over an entire 2-sphere worth of measurements, and they are valid \emph{only} at leading order in the asymptotically flat expansion. Of course, if one is interested in approximation and not exact expressions, then one could envisage using a sphere small enough to be accessible experimentally, while being large enough for the curvature effects of the gravitational system inside to be small enough, all the while neglecting the influence from the outside gravitational system, its gravitational waves and potentials affecting the interior and the back-reaction of the experiment. One would need to be able to control all of these effects in order to estimate the corrections to the formula and know the validity of \linebreak the {approximation.} Only then will the procedure be applicable to real experiments.

\section{Discussion}
\label{s4}

The total energy-momentum $\admP$, and the 4-momentum $\bsP$ that is left over after allowing for radiation to escape to infinity until a retarded instant of time, are  basic observables of fundamental importance for isolated systems in general relativity {\smash{\cite{adm,bondi,sachs}.}} But, as emphasized in section \ref{s1}, they are also surprisingly subtle. First, since the gravitational field itself contributes to energy momentum in general relativity, the ADM 4-momentum $\admP$ is a \emph{global} concept, defined at spatial infinity. This is in striking contrast to Newtonian gravity, where one can extract the total mass by calculating the flux of the gradient of the Newtonian potential across any finite 2-sphere outside the sources. Second, even when matter sources --e.g. a Maxwell field-- extend all the way to infinity, $\admP$ and $\bsP$ can be expressed as \emph{2-sphere integrals},\,  and the integrand consists \emph{only} of gravitational fields; matter fields do not appear.

Being 2-sphere integrals, $\bsP$ and $\admP$ are \emph{global in angles}. For the time-dependent  Bondi-Sachs 4-momentum, this is not surprising because one can envisage gravitational radiation beamed only in a small solid angle. If this occurs, it is intuitively clear that knowledge of fields in the complement of that solid angle will not suffice to determine the $\bsP$ that is left over after the passage of the burst of that gravitational wave. But the global nature in angles is less intuitive for $\admP$. Indeed, in stationary space-times one routinely determines the ADM  mass from orbits of  satellites around the central body that sense the gravitational field along curves that constitute only a set of zero measure in the 2-sphere spanned by all angles. But as the Carlotto-Schoen gluing construction \cite{carlotto-schoen} vividly demonstrates, this cannot be done more generally because orbits can lie in a flat region of space-time even when the total ADM mass is non-zero. These subtle features could not have been foreseen before the advent of global methods in general relativity  because they go against intuition derived from properties of energy-momentum in non-gravitational field theories as well as Newtonian gravity. 

It is not surprising, then, that expressions of $\admP$ and $\bsP$ were first obtained by mathematical considerations rather than physical expectations about the notion of energy-momentum. $\admP$ was first introduced using Hamiltonian methods \cite{adm,rt},\, and $\bsP$ from the asymptotic behavior of solutions to Einstein's equations and Bianchi identities that provided useful mathematical balance laws between certain 2-sphere `charge-integrals' and `3-surface fluxes' \cite{bondi,sachs}. Confidence in the physical interpretation of the 2-sphere charge integrals was significantly enhanced by the positive energy theorems \cite{sy,ew,ghmp,sy2,rt}\, in both regimes. However, from their expressions (\ref{ADMenergy}) and (\ref{BSE1}) themselves, it is far from being obvious \emph{why} these 2-sphere integrals should have the physical connotations of energy. 

Therefore, in section \ref{s2.2} we first recast the expression (\ref{ADMenergy}) of the ADM energy in terms of 4-dimensional curvature using Einstein's equations, and showed that it is directly related to the \emph{tidal acceleration} of test particles (spread on two nearby concentric spheres) caused by the space-time curvature produced by a massive central body. While the nature of gravity undergoes a profound transformation in the passage from Newtonian gravity to general relativity, the notion of tidal acceleration remains in tact. Thus, one can say that tidal accelerations \emph{capture the very essence of what the total gravitational energy does}. In section \ref{s2.3} we provided a thought experiment to measure the ADM energy using tidal acceleration, thereby providing an operational meaning of $E_{\rm ADM}$. This experiment was based on the expression (\ref{riem-exp}) of the ADM energy in terms of the 4-d Riemann tensor. To arrive at this expression, we recast $E_{\rm ADM}$ as a 2-sphere integral (\ref{weyl-exp}) involving the electric part of the Weyl tensor, and the integral (\ref{ricci-exp2}) in terms of the 3-d Ricci tensor. These expressions are of interest in their own right. For example, as we already noted, the Carlotto-Schoen analysis begins with (\ref{ricci-exp2}),\, and the expression (\ref{weyl-exp}) is used in relating the ADM and Bondi-Sachs 4-momenta in space-times that are asymptotically flat in both regimes \cite{ak1}. It also provides us with 4-momenta at $\scri$ in presence of a negative cosmological constant \cite{aaam-ads}.

In section \ref{s3} we turned to the Bondi-Sachs energy $E_{\rm BS}$. In absence of gravitational waves, $E_{\rm BS}$ is conserved and one can again use the thought experiment of section \ref{s2} to measure it \cite{aasb}. But in presence of gravitational waves, this strategy fails to capture the full content of $E_{\rm BS}$. Therefore, following the same strategy as in section \ref{s2},\, in section \ref{s3.2} we first recast the expression (\ref{BSE1}) of $E_{\rm BS}$ in a form (\ref{BSE4}) that directly captures a physical effect induced by $E_{\rm BS}$: it dictates the (angular average of the) convergence $|\theta_{(n)}|$ of ingoing null rays in the asymptotic region. In Section \ref{s3.3} we  provided a protocol to determine the angular average of $|\theta_{(n)}|$ by measuring areas of suitably chosen successive converging light fronts. The result is a thought experiment to measure $E_{\rm BS}$. It is rather remarkable that while the expression (\ref{BSE4}) depends \emph{only on} $\theta_{(n)}$, it manages to capture not only the\, `Coulombic part'\, of the Weyl tensor (the first term in the integrand of (\ref{BSE1})), but also on the precise radiative term that enters the expression of the Bondi-Sachs energy (the second term in (\ref{BSE1})), \emph{and nothing else}! From a physical standpoint, then, the Bondi-Sachs energy $E_{\rm BS}$ manifests itself in the excess in the convergence of ingoing light rays, over and above that which is  already present in Minkowski space-time. One can take this as the defining property of the\, `left-over energy at a retarded instant of time'.\, The fact that this property refers only to space-time geometry \,`explains'\, the absence of matter fields in the expression of $E_{\rm BS}$. 

In section \ref{s1} we emphasized that it is hard to see why the original expressions (\ref{ADMenergy}) and (\ref{BSE1}) of $E_{\rm ADM}$ and $E_{\rm BS}$ have the connotations of energy. But these are the expressions used, e.g., in the LIGO-Virgo analyses that provide us with the parameters that label the compact binaries, including the initial ADM masses of the progenitors and the final Bondi-Sachs mass of the remnant. Thus, there is a clear sense in which $\admP$ and $\bsP$ have been measured for compact coalescing binaries. But these measurements do not shed any light on the\, `operational meaning'\, of $E_{\rm ADM}$ and $E_{\rm BS}$ because these analyses begin by \emph{assuming} that (\ref{ADMenergy}) and (\ref{BSE1}) have the interpretation that is assigned to them; the question of \emph{why} they have this connotation is not an issue they are concerned with. For their purposes, it suffices to just assume that they do. Put differently, the expressions (\ref{ADMenergy}) and (\ref{BSE1}) provide parameters that serve as labels along with other  \,`intrinsic parameters'\,  --the spins of progenitors, eccentricity and orientation of their orbit-- as well as 4\, `extrinsic  parameters' \, that characterize the location of the source with respect to the detector. For each set of values for these 10 intrinsic and 4 extrinsic parameters, the waveform models provide the time-dependent strain that detectors should see. Comparing these theoretical waveforms with the observed strains, one finds the best fit values for the entire set of parameters (as well as the corresponding posterior probability distributions). Thus, one does not directly measure the masses. The goal of the LIGO-Virgo analysis is to find the \emph{joint} best-fit values (together with error bars) since this is the output that characterizes the binary and provides us with the detailed picture of its coalescence.  
In this paper, the focus is on a narrow but rather fundamental issue as to \emph{why} the expressions in (\ref{ADMenergy}) and (\ref{BSE1}) have the connotations they carry. Our thought experiment  provides a direct measurement of these quantities and, as a consequence, explains why they have this connotation. Our goal is different and much more modest.

Finally, it is our hope that the expression (\ref{BSE4}) of $E_{\rm BS}$ in terms of $\theta_{(n)}$ will be useful in future investigations especially in non-perturbative quantum gravity. We will conclude with a summary of the main ideas. Recall  first that $\scripm$ provide a natural arena to construct Hilbert spaces of states to investigate gravitational scattering in a \emph{non-perturbative} setting of the so-called \,`Asymptotic Quantization Program'\, (AQP) \cite{aa-biblio}. In recent years these asymptotic Hilbert spaces have been used very effectively to analyze the relation between infrared issues in the AQP and Weinberg's\, `soft theorems'\, in perturbative quantum gravity (see \cite{Strominger} for a summary). However, so far, the problem of defining the quantum operator $\hat{E}_{\rm BS}[S_\circ]$ corresponding to Bondi-Sachs energy has remained open. The essential difficulty has been that the original AQP and the subsequent work is focused on the radiative aspects of the gravitational field, while the standard expression (\ref{BSE1}) of $E_{\rm BS}$ involves a Coulombic field as well. And there is no operator on the asymptotic Hilbert spaces representing these fields. (This is a deeper conceptual problem, in addition to the mathematical issues associated with regularization of products of operator-valued distributions involving radiative fields.) 

The ongoing work on null surfaces in loop quantum gravity (LQG) has opened exciting possibilities for analyzing various non-perturbative features of dynamics (see, e.g., \cite{Speziale:2013ifa,Wieland:2017cmf,DePaoli:2017sar,ww1,fops,fpr,fgw,ww2,cfl}). In particular, when combined with our expression (\ref{BSE4}) of $E_{\rm BS}$, they provide a path to define the Bondi-Sachs energy operator $\h{E}_{\rm BS}[S_\circ]$. For, there are well-defined operators on the LQG Hilbert space\, --in particular the area operator-- \,that capture \,`non-radiative'\, degrees of freedom and therefore have no counterparts in Hilbert spaces used in perturbative treatments, nor those that feature in the AQP. Since the only dynamical variable that features in the our expression (\ref{BSE4}) of $E_{\rm BS}$ is  $\theta_{(n)}$,\, and since it refers to the rate of change of the areas of  2-spheres, one can hope to define the operator $\h{E}_{\rm BS}$ on the LQG Hilbert space. This task would be facilitated by restricting oneself to asymptotically flat space-times at the outset, and  getting rid of `gauge diffeomorphisms'\, --i.e., those, which are asymptotically identity-- \, by fixing the equivalence class of Bondi-Sachs charts discussed in section \ref{s3}, where any two are related by a BMS diffeomorphism, i.e., an asymptotic symmetry. One would introduce area operators associated with 2-spheres $u=u_\circ$ and $r= r_\circ$ and define the required operator $\h{E}_{\rm BS}$ using the operator analog of Eq. (\ref{BSE5}). 

Presumably $\h{E}_{\rm BS}$ would have a discrete spectrum because area operators do. And this discreetness may play an important role in the analysis of black hole evaporation. For example,  the analysis of the (classically) exactly soluble Callen, Giddings, Harvey, Strominger (CGHS) model in the mean field approximation
\footnote{In this approximation one ignores the quantum fluctuations of geometry --i.e., the geometrical operators are replaced by their expectation value-- but retains those in matter. This is justified if we have a large number of matter fields (i.e. evaporation channels) all coupled to a single gravitational field. The universality discussed below refers to the ADM and Bondi-Sachs masses \emph{per evaporation channel}.} 
shows that there is a direct correlation between $E_{\rm BS}$ and the area of the dynamical horizon DH that evaporates {\smash{\cite{apr2}.} Since the area of the DH would be quantized in LQG, it stands to reason that  $E_{\rm BS}$ would also be quantized. Evaporation of the DH could then be interpreted in detail as a process in which quanta of area of the DH are converted to energy quanta at infinity. The evaporation process of the CGHS black hole has a rather mysterious feature that irrespective of how large the initial or ADM mass is, the Bondi-Sachs mass of the black hole at the end of the semi-classical phase (beyond which the mean field approximation cannot be trusted) has a universal value. This could perhaps be\, `explained'\, in LQG using the fact that there is an area-gap. For 4-d black holes of direct physical interest, one would like to study correlations between observables at the evaporating DH and those at $\scrip$. Availability of operators corresponding to the area of the DH and $\h{E}_{\rm BS}$ would be very helpful in this respect. Availability of $\h{E}_{\rm BS}$ would be useful also in contexts that do not involve black holes, e.g. in the discussion of a possible  \,`holographic nature'\, of $\scrip$ (see, in particular, \cite{lprs}). These analyses often assume that there is a well-defined operator $\h{E}_{\rm BS}$ with certain properties. It would be interesting to check whether or not these assumptions are satisfied by the LQG operator, defined along the lines sketched above.

\section*{Acknowledgments}

This work was supported in part by Eberly and Atherton Research Funds and by the Distinguished Visiting Research Program of the Perimeter Institute.


\begin{thebibliography}{99}

\bibitem{carlotto-schoen} A.~Carlotto and R.~Schoen, Localizing solutions of the Einstein constraint
equations, Invent. Math. \textbf{205}, 559-615 (2016); arXiv:1407.4766 [math.AP].

\bibitem{schoen} R. Schoen, Localizing solutions of the Einstein Equation,\\ philippelefloch.org/wp-content/uploads/2015/11/2015-ihp-richardschoen.pdf

\bibitem{pc-bourbaki} P.~T.~Chrisciel, Anti-gravity \`a la Carlotto-Schoen,\\ http://www.bourbaki.ens.fr/TEXTES/1120.pdf

\bibitem{adm} R.~Arnowitt, S.~Deser and C.~W.~Misner, The dynamics of general relativity, In: \textit{Gravitation: An Introduction to Current Research}, edited by L. Witten (Wiley \& Sons, New York, 1962).
    
\bibitem{rt} T.~Regge and C.~Teitelboim, Role of surface integrals in the Hamiltonian formulation of general relativity,
Ann. Phys. \textbf{88}, 286 (1974).

\bibitem{komar} A.~Komar. Covariant conservation laws in general relativity, Phys. Rev. \textbf{113}, 934-36 (1959).

\bibitem{beig} R.~Beig, Arnowitt-Deser-Misner energy and $g_{00}$, Phys. Lett. A\textbf{69}, 153-155 (1978).

\bibitem{aaam-conserved} A. ~Ashtekar and A.~Magnon, On conserved quantities in general relativity, J. Math, Physics, \textbf{20} 793-800 (1979).

\bibitem{aaam3+1}A.~Ashtekar and A.~Magnon, From $\inot$ to the 3+1 decomposition of spatial infinity, J. Math. Phys. \textbf{25}, 2682-2690 (1984).


\bibitem{bondi} H.~Bondi, M.~van der Burg, and A.~Metzner, Gravitational waves in genera1 relativity VII. Waves from axisymmetric isolated systems, Proc. R. Soc. (London) A\textbf{269}, 21 (1962).

\bibitem{sachs} R.~K.~Sachs, Gravitational waves in general relativity VIII. Waves in asymptotically flat space-times Proc. R. Soc. (London) A\textbf{270}, 103 (1962).

\bibitem{aams} A.~Ashtekar and M.~Streubel, Symplectic geometry of
    radiative modes and conserved quantities at null infinity, Proc. R. Soc. (London) \textbf{A376}, 585-607 (1981).


\bibitem{aass2}A.~Ashtekar and S. Speziale, Null Infinity and Horizons: A New Approach to Fluxes and Charges, Phys. Rev. D \textbf{110} (2024) no.4, 044049.

\bibitem{sy}R.~Schoen and S.~T.~Yau, Proof of the positive energy theorem.II, Commun. Math.Phys. \textbf{79}, 231-260 (1981).

\bibitem{ew} E.~Witten, A new proof of the positive energy theorem, Commun. Math. Phys. \textbf{80}, 381 (1981).

\bibitem{ghmp} Gary~T.~Horowitz and Malcolm~J.~Perry, Gravitational energy cannot become negative, Phys. Rev. Lett. \textbf{48}, 371- 374 (1982)

\bibitem{sy2} R.Schoen and S.~T.~Yau, Proof that the Bondi mass is positive, Phys. Rev. Lett. 48, 369-370 (1982).  

\bibitem{orpt} O.~Reula and K.~P.~Tod, Positivity of Bondi energy, J. Math. Phys. \textbf{25}, 1004-1008 (1984).

\bibitem{aasb} A.~Ashtekar and S.~Baharami, Asymptotics with a positive cosmological constant. IV. The no-incoming radiation condition, Phys. Rev. D\textbf{100}, 024042 (2019).

\bibitem{szabados} L.~B.~Szabados, Quasi-local energy-momentum and angular momentum in GR: A Review article, Liv. Rev. (Relativity), \textbf{7}, 4 (2004).


\bibitem{aa-ein} A.~Ashtekar, Asymptotic Structure of the
    Gravitational Field at Spatial Infinity, In: \textit{General
    Relativity and Gravitation: One Hundred Years After the birth of
    Albert Einstein,} edited by A. Held, (Plenum, New York, 1980).

\bibitem{etnrp} R.~Penrose and E.~T.~Newman, New conservation laws for zero rest mass fields in asymptotically flat space-times, Proc. R. Soc. (London) A\textbf{305}, 175-204 (1968). 

\bibitem{rp} R.~Penrose, Zero rest mass fields including gravitation: asymptotic behavior, Proc. R. Soc. (London)
    A\textbf{284}, 157-203 (1965).
    
\bibitem{swh} S.~W.~Hawking, Gravitational radiation in an expanding universe, J. Math. Phys. \textbf{9} 598-604 (1968).

\bibitem{aaam-prl} A.~Ashtekar and A.~Magnon, Energy-momentum in general relativity, Phys. Rev. Lett. \textbf{43}, 181-184 (1979).

\bibitem{ak1} A.~Ashtekar and N.~Khera, Unified treatment of null and spatial infinity III: Asymptotically Minkowski space-times, JHEP \textbf{02}, 210 (2024).


\bibitem{aaam-ads} A.~Ashtekar and A.~Magnon, Asymptotically anti-de Sitter Space-times, Class. Quant. Grav. Lett. \textbf{1}, L39-L41 (1984).

\bibitem{aa-biblio} A.~Ashtekar, \emph{Asymptotic quantization: based on 1984 Naples lectures}, Bibliopolis, Naples
(1987)

\bibitem{Strominger} A.~Strominger, \emph{Lectures on the infrared structure of gravity and gauge theory} (Princeton University Press, Princeton 2018).

\bibitem{Speziale:2013ifa}
S.~Speziale and M.~Zhang, Null twisted geometries, Phys. Rev. D \textbf{89} (2014) no.8, 084070.

\bibitem{Wieland:2017cmf}
W.~Wieland, Fock representation of gravitational boundary modes and the discreteness of the area spectrum,
Annales Henri Poincare \textbf{18} (2017) no.11, 3695-3717.

\bibitem{DePaoli:2017sar}
E.~De Paoli and S.~Speziale,
Sachs\textquoteright{} free data in real connection variables, JHEP \textbf{11} (2017), 205. 

\bibitem{ww1} W.~Wieland, Null infinity as an open Hamiltonian system, JHEP \textbf{04}, 095 (2021).

\bibitem{fops} L.~Freidel, R.~Oliveri, D.~Pranzetti and S.~Speziale, Extended corner symmetry, charge
bracket and Einstein's equations, JHEP \textbf{09}, 083 (2021).

\bibitem{fpr} L.~Freidel, D.~Pranzetti and A.-M.~Raclariu, Higher spin dynamics in gravity and $w_{1+\infty}$ 
celestial symmetries, Phys. Rev. D\textbf{106},  086013 (2022).

\bibitem{fgw} L.~Freidel, M. Geiller, W.~Wieland, Corner symmetry and quantum geometry, arXiv \texttt{2302.12799}.

\bibitem{ww2} W.~Wieland, Quantum geometry of the null cone, arXiv \texttt{2401.17491}. 

\bibitem{cfl} L.~Ciambelli, L.~Freidel, R.~G.~Leigh, Quantum null geometry and gravity, \texttt{2407.11132}.


\bibitem{apr2} A.~Ashtekar, F.~Pretorius and F.~Ramazanoglu, Evaporation of two dimensional black holes, Phys. Rev.D\textbf{83}, 044040 1-18 (2011).

\bibitem{lprs} A.~Laddha, S.~G.~Prabhu, S.~Raju and P.~Shrivastava, The Holographic Nature of Null Infinity,  SciPost Phys. \textbf{10}, 041 (2021). 

\end{thebibliography}
\end{document}